\documentclass[11pt,twoside]{article}

\usepackage{galev06}
\usepackage{epsf}
\usepackage{psfig}
\usepackage{lscape}

\begin{document}
\title{SED of starbursts}   
\author{Ralf Siebenmorgen$^1$ and Endrik Kr\"ugel$^2$}   
\affil{$^1$ European Southern Observatory, Karl-Schwarzschildstr. 2,
        D-85748 Garching b. M\"unchen, Germany \newline
$^2$ Max-Planck-Institut f\"ur Radioastronomie, Auf dem H\"ugel 69,
        Postfach 2024, D-53010 Bonn, Germany}

\begin{abstract} 
We provide a library of some 7000 SEDs for starbursts and ultra
luminous galaxies ({\tt
http://www.eso.org/\~\/rsiebenm/sb\_models}). Its purpose is to
quickly obtain estimates of the basic parameters, such as luminosity,
size and dust or gas mass and to predict the flux at yet unobserved
wavelengths.  The procedure is simple and consists of finding an
element in the library that matches the observations.  The objects may
be in the local universe or at high $z$.  We calculate the radiative
transfer in spherical symmetry for a stellar cluster permeated by an
interstellar medium with standard (Milky Way) dust properties.  The
cluster contains two stellar populations: old bulge stars and OB
stars.  Because the latter are young, a certain fraction of them will
be embedded in compact clouds which constitute hot spots that
determine the MIR fluxes.  We present SEDs for a broad range of
luminosities, sizes and obscurations.  We argue that the assumption of
spherical symmetry and the neglect of clumpiness of the medium are not
severe shortcomings for computing the dust emission.  The validity of
the approach is demonstrated by matching the SED of the best studied
galaxies, including M82 and Arp220, by library elements; one example
is shown for a galaxy at high redshift ($z \sim 3$). Generally, one
finds an element which fits the observed SED very well, and the
parameters defining the element are in full accord with what is known
about the galaxy from detailed studies.

\end{abstract}


\noindent {\bf Introduction:} \/ By definition, the rapid conversion of 
a large amount of gas into predominantly massive ($> 8 M_\odot$)
stars, or the result of such a conversion, is called a starburst.
Starburst galaxies constitute a unique class of extragalactic objects.
The phenomenon is of fundamental importance to the state and evolution
of the universe (Heckman 1998). Starbursts are also cosmologically
significant if one interprets the high bolometric luminosities of high
redshift galaxies to be due to star formation (Chary \& Elbaz
2001). To interpret infrared observations and to arrive at a
self--consistent picture for the spatial distribution of stars and
interstellar matter in the starburst nucleus and of the range of dust
temperatures, one has to simulate the transfer of continuum radiation
in a dusty medium (e.g. Kr\"ugel \& Siebenmorgen 1994).  Line emission
is energetically negligible.

\noindent
A starburst has four basic parameters: total luminosity, $L$, dust or
gas mass, $M_{\rm d}$ or $M_{\rm gas}$, visual extinction, $A_{\rm
V}$, and size.  Size, $A_{\rm V}$ and $M_{\rm d}$ are, of course,
related, for a homogeneous density model, only two of them are
independent. The luminosity follows observationally in a straight
forward way by integrating the spectral energy distribution over
frequency, $M_{\rm d}$ is readily derived from submillimeter data, and
the outcome is almost independent of the internal structure or viewing
angle of the starburst.  The size is best obtained from radio
observations.

\begin{figure}[!ht]
\centerline{\psfig{figure=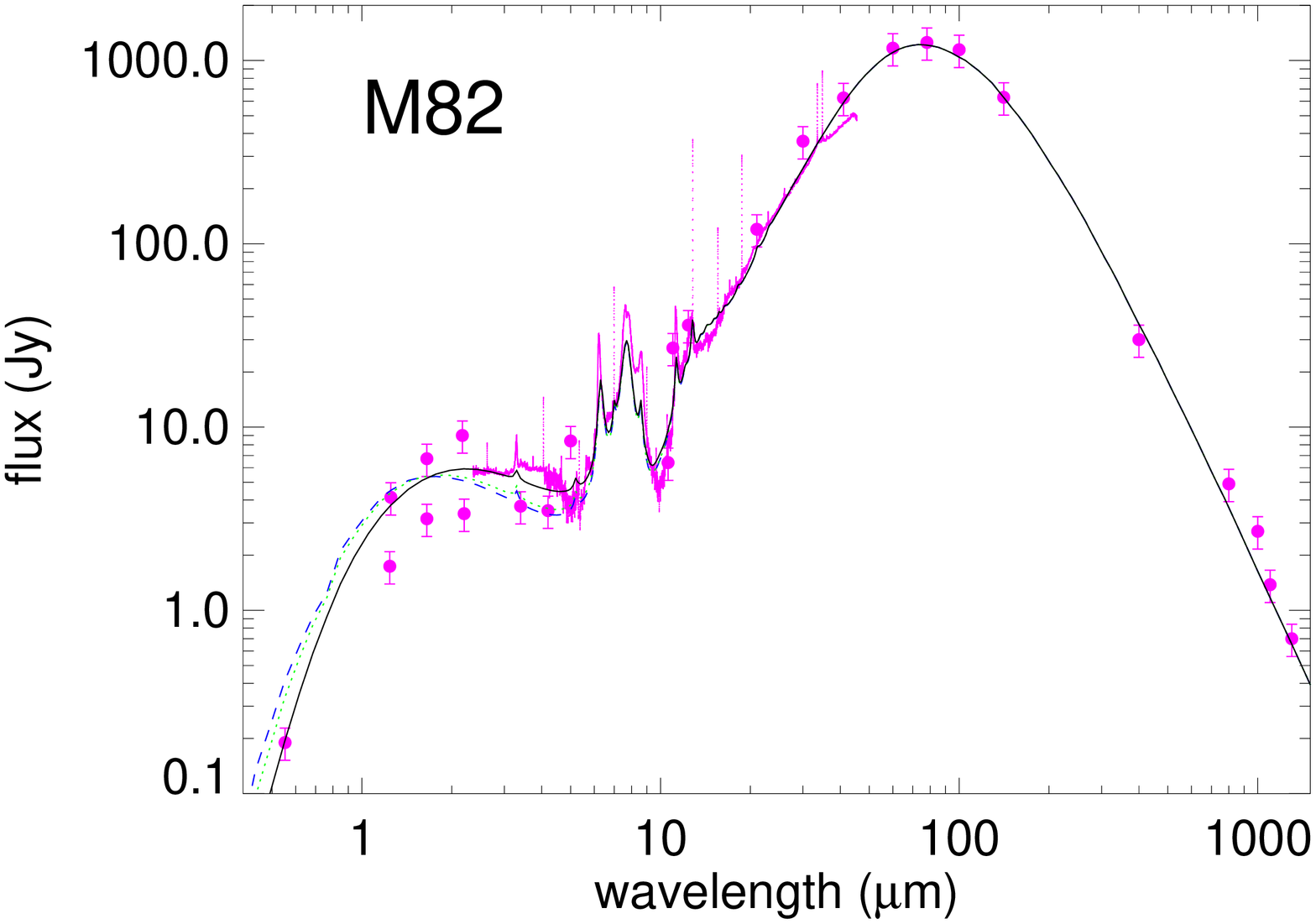,angle=90,width=\textwidth}} 
\centerline{\psfig{figure=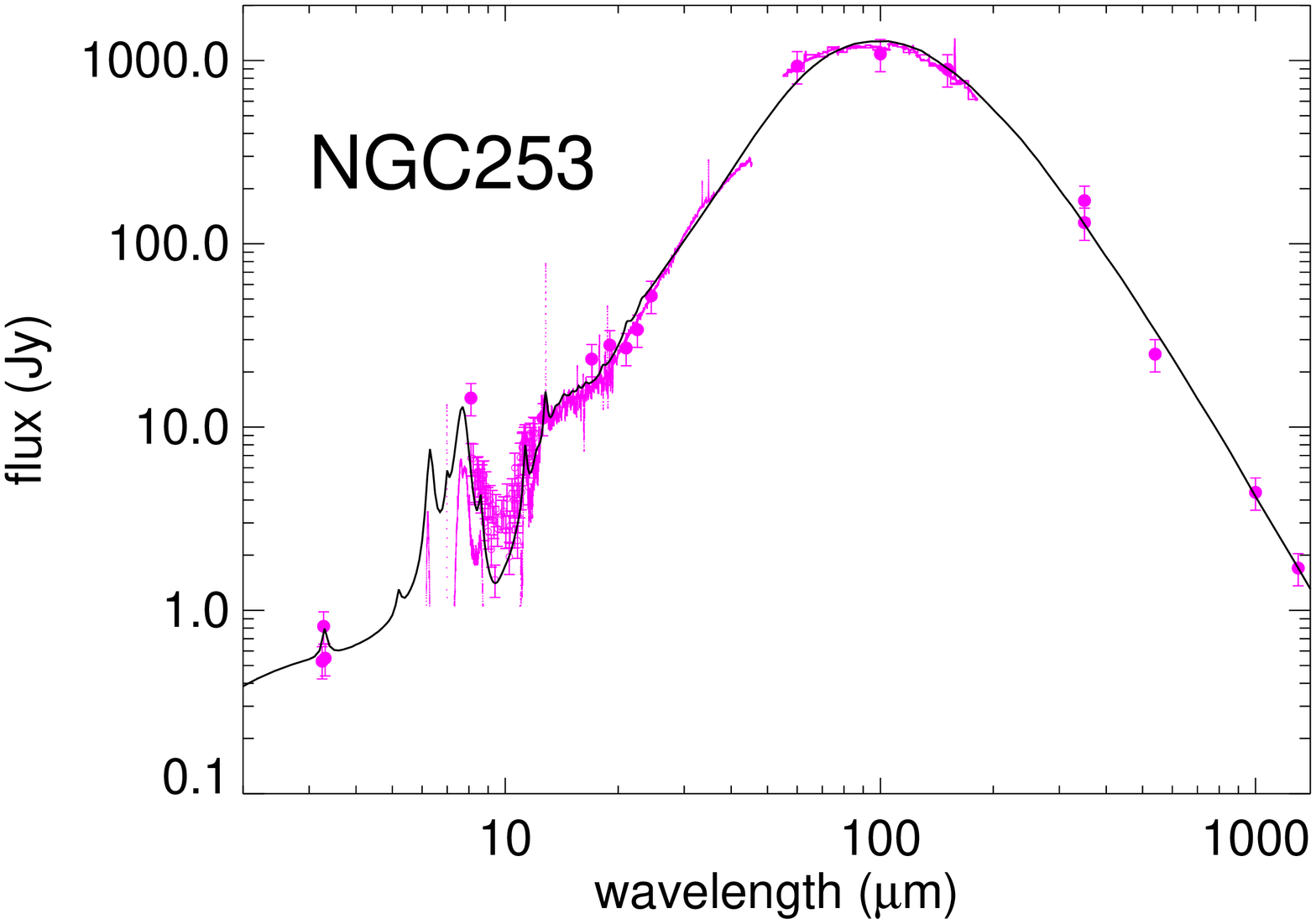,angle=90,width=\textwidth}} 
\caption{SED of the central region of M82 ({\itshape top}) and NGC253
({\itshape bottom}). Data points with $1\sigma$ error bar as available
in NED, ISO, IRAS; NIR photometry in 40$''$ -- 100$''$
aperture.  Full line: library model with parameters for M82
($L^{\rm{tot}}=10^{10.5} L_\odot$, $D=3.5$Mpc, $R=350$pc, $A_{\rm
V}=36$mag, $L_{\rm {OB}}/L^{\rm {tot}} =0.4$, $n^{\rm
{hs}}=10^4$cm$^{-3}$) and NGC253 ($L^{\rm{tot}}=10^{10.1} L_\odot$,
$D=2.5$Mpc, $R=350$pc, $A_{\rm V}=72$mag, $L_{\rm {OB}}/L^{\rm {tot}}
=0.4$, $n^{\rm {hs}}=7500$cm$^{-3}$). For M82 to fit data below
5$\mu$m, we added to the SED library spectrum a blackbody, either
unreddened ($T= 2500$ K, full line), or reddened ($T=8000$K, $A_{\rm
V} =4$mag, dashed, or $T=5000$K, $A_{\rm V} =3$mag, dotted).}
\end{figure}

\clearpage

\noindent {\bf Model description:} \/ We use standard dust.  It consists of
large silicate and amorphous carbon grains with a size distribution.
A population of small graphite grains ($<100$\AA) and two kinds of
polycyclic aromatic hydrocarbons (PAH). Our dust mixture produces
reddening in rough agreement with the standard interstellar extinction
curve for $R_{\rm V} = 3.1$. An important feature of our radiative
transfer model is the division of the sources in the starburst nucleus
into two classes. {\it a)} OB stars in dense clouds with total
luminosity $L_{\rm OB}$.  The immediate surroundings of such a star
constitutes a {\it hot spot} which determine the MIR part of the SED
of the nucleus. {\it b)} The total luminosity of all other stars is
$L_{\rm tot}-L_{\rm OB}$. They are not enveloped in a dense cloud. For
detailed description see Siebenmorgen \& Kr\"ugel (2006). In the
following we give three examples.

\begin{figure}[!ht]
\centerline{\psfig{figure=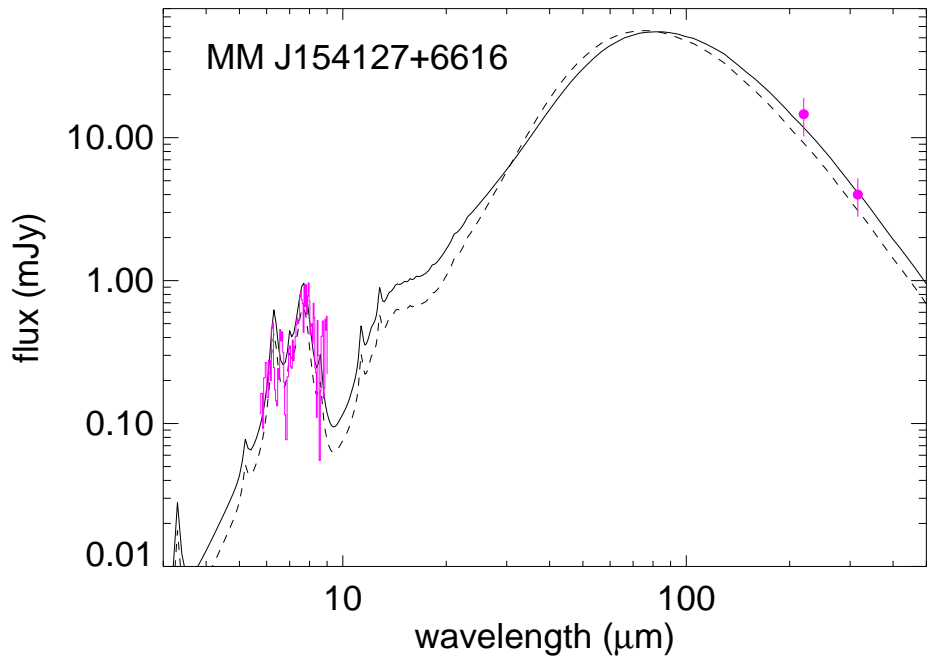,angle=90,width=\textwidth}} 
\caption{The spectrum of MMJ154127+6616 in the restframe of the
   galaxy ($z=2.8$) with two library models. Model parameters: {\itshape 
   solid line:} $L= 10^{13.7}$ L$_\odot$, \, A$_{\rm V}=70$ mag, $R=$
   15kpc; {\itshape  dashed line:} $L=10^{13.7}$ L$_\odot$, \, A$_{\rm
   V}=120$ mag, $R=$ 9kpc.  Observational points from Lutz et
   al.~(2005), Eales et al.~(2003) and Bertoldi et al.~(2000).}
\end{figure}

\noindent {\bf M82:} \/ The present model for this archetype starburst is
similar to the one proposed before (Kr\"ugel \& Siebenmorgen 1994). At
shorter wavelengths, the observed flux does not steeply decline, as
one would expect judging from the depth of the silicate absorption
feature ($A_{\rm V} \ge 15$ mag).  Therefore, either hard radiation
leaks out because of clumps or funnels created by supernova
explosions, or there are stars in M82 outside the opaque nuclear dust
clouds.  As our model cannot handle clumping, but we nevertheless wish
to extend the spectrum into the UV, we simply add another stellar
component, which is not included in a self--consistent way, but as its
luminosity is $\sim 10\%$ of the total, such an approximation may be
tolerable.  The stellar temperature and foreground reddening of the
additional component are poorly constrained (Fig.1).  This is also
reflected by the controversial interpretations via an old stellar
population (Silva et al. ~1998) or via young, but obscured stars
(Efstathiou et al.~2000).  \\

\noindent {\bf NGC253:} shows at high resolution in the MIR a complex 
structure.  Nevertheless, the low spatial resolution observations are
well reproduced in our fit (Fig.1). Below 2 Jy, ISOSWS data are noisy
and have therefore been omitted.  The dip at 18$\mu$m in the model of
Piovan et al.~(2006) is not present in ours, and not borne out by the
observations. \\

\noindent {\bf High redshift galaxies:} \/ The library also seems
to be applicable to extremely luminous objects at large redshifts.  We
demonstrate this in Fig.2 for the submillimeter galaxy MMJ154127+6616
which is at $z=2.8$ as derived from the identification of PAH features
in the Spitzer IRS low resolution spectrum (Lutz et al., 2005).  It
seems to be an extremely massive object ($10^{11}$ M$_\odot$) where
the starburst comprises the whole galaxy.  For luminosities above
$10^{12.7}$ L$_\odot$, we have therefore modified the structure of the
models accordingly.  {\it a)} The size is increased to $R=9$ kpc and
15 kpc.  {\it b)} There are only OB stars as giants have had no time
to evolve.  40\% of the stars are assumed to be in regions of enhaned
density (hot spots).  {\it c)} The star formation region extends to
the edge of the galaxy. One finds two entries in the library which are
compatible with the observations (Fig.2), both imply
extreme luminosities (L$_{\rm dust} = 10^{13.7}$ L$_\odot$) and dust
masses close to $10^{10}$ M$_\odot$. \\

\noindent {\bf Conclusion:} \/ A set of spectral energy distributions for 
starbursts covering a wide range of parameters is presented by
Siebenmorgen \& Kr\"ugel (2006). Anyone with
infrared data and interested in their interpretation can compare them
with our models, find an SED that matches (after normalization of the
distance) and thus constrain the properties of the starburst under
investigation without having to perform a radiative transfer
computation himself. \\

\noindent {\bf Acknowledgements:} \/ We thank D. Lutz for providing the 
observed SED of MM J154127+6616. 



\end{document}